\documentclass[12pt]{article}
\pdfoutput=1
\usepackage{epsfig}
\usepackage{amsfonts}
\usepackage{amssymb}
\usepackage{xcolor}
\usepackage{epsfig}
\usepackage{amsfonts}
\usepackage{amssymb}
\usepackage{xcolor}
\usepackage{mathrsfs}
\usepackage{amsmath}
\usepackage{upgreek}
\usepackage{graphicx}
\usepackage{cite}
\usepackage{calligra}
\usepackage{caption}
\usepackage{subcaption}
\usepackage{cite}
\usepackage{calligra}
\usepackage[english]{babel} 
\usepackage{hyperref}

\begin{document}
\newcommand{\nc}{\newcommand}
\nc{\beq}{\begin{equation}} \nc{\eeq}{\end{equation}}
\nc{\beqa}{\begin{eqnarray}} \nc{\eeqa}{\end{eqnarray}}
\nc{\R}{{\cal R}}
\nc{\A}{{\cal A}}
\nc{\K}{{\cal K}}
\nc{\B}{{\cal B}}
\begin{center}

{\bf \Large  New Solutions for RG Equations in QCD} \vspace{1.0cm}

{\bf \large R.M. Iakhibbaev$^{1,a}$, D. I. Kazakov$^{1,b}$ and D.M.Tolkachev$^{1,2,c}$} \vspace{0.5cm}

{\it $^1$Bogoliubov Laboratory of Theoretical Physics, Joint Institute for Nuclear Research,
  6, Joliot Curie, 141980 Dubna, Russia \\and \\
$^2$Stepanov Institute of Physics,
68, Nezavisimosti Ave., 220072, Minsk, Belarus}
\vspace{0.5cm}

\abstract{We construct simple analytical solutions of renormalization group equations for the running coupling and for the Green functions in QCD in the asymptotic regime. These solutions have an explicit form and subsequently sum up the leading, subleading, and so on logarithms  in all orders of PT.  They easily reproduce the  inverse logarithm expansion and  allow for further summation and improvement of the asymptotic behaviour.}
\end{center}

\text{\footnotesize{ E-mails: $^a$yaxibbaev@jinr.ru, $^b$kazakovd@theor.jinr.ru, $^c$dtolkachev@jinr.ru,
}}

\text{\footnotesize{Keywords: RG equations, Running coupling, Approximate analytic solution to the RGE}}

 \section{Introduction}
 
 It is impossible to imagine the application of QCD perturbative expansion without improving the renormalisation group (RG). The notion of the running coupling is the key ingredient of this analysis of PT series \cite{BogoliubovBook,Collins}.  At the same time, the RG equations  for the coupling, Green functions, structure functions, and so on, being exact,  are written in a given order of PT.  This means that the beta-function and the appropriate anomalous dimensions are calculated in PT with  limited accuracy. Then, one has to solve these approximate differential equations, and one usually tries to solve them exactly. Needless to say, these solutions have a simple analytical form only in the first one-loop approximation. Starting with two loops, one either has a solution in terms of the Lambert $W$-function \cite{corless1993lambert} (for the two loop coupling constant), or one has no analytic solution at all and has to invent approximate expansions in different regimes \cite{Shirkov:1981mb,Magradze:1999um}. Apparently, the analytic solution is preferable since it allows one to analyse its behaviour and is easy for numerical programming.
 
In these circumstances, we notice that the applied strategy, namely to write down the RG equations in an approximate form in a given order of PT and to solve these approximate equations exactly, is not adequate. 
In fact, the aim of improving the PT series by using the RG  equations is to sum up the leading asymptotics in all orders of PT. Thus, the solution of the one-loop RG equation sums up the leading logarithms. One can say that exact solution of the two-loop equation sums up all the next-to-leading (NL) logs, and so on. However, this is not the case. The exact solution of the two-loop equation indeed sums the NL logs but it also sums part of the NNL logs. Further on, the exact solution of the three-loop equation (which is hardly possible) will sum the NNL logs and part of the NNNL logs, and so on.

Surprisingly, if one wants to present the strategy that allows one to sum up the leading, NL, NNL, etc logs
at each step without any mixture with the lower terms, one gets  a relatively simple scheme which allows for the explicit analytical form at each stage.  In this note, we present this type of scheme for solving the RG equations for the running coupling and the functions with the  anomalous dimensions in the case of a single coupling theory like QCD. The explicit formulas for improvement of the logarithmic expansion are provided.

\section{Perturbative Expansion and New Solutions}
  
 In a theory with a single coupling like QCD the running coupling obeys the RG equation
\beq
\frac{d}{dL} \bar\alpha[\alpha,L]=\beta(\bar \alpha)=\beta_0 \bar\alpha^2+\beta_1 \bar\alpha^3+\beta_2 \bar\alpha^4+\beta_3 \bar\alpha^5+\dots,  \ \ L\equiv \log Q^2/\mu^2  \label{eq}
\eeq
with the boundary condition $\bar \alpha [\alpha,0]=\alpha$,
where the beta-function is calculated  within perturbation theory up to a given order of PT.

The formal solution of eq.(\ref{eq}) in quadratures has the form
\beq
\log(Q^2/\mu^2)=\int_{\alpha}^{\bar\alpha} \frac{dx}{\beta(x)}= F[\bar\alpha]-F[\alpha]
\eeq
It can  also be rewritten as
\beq
\log(Q^2/\Lambda^2)= F[\bar\alpha],
\eeq
if one expresses the solution in terms of the one-dimensionfull variable  $\Lambda$ related to the original normalization parameter $\mu$ by the condition
\beq
\Lambda^2 =\mu^2 \exp(-F[\alpha]) .\label{lam}
\eeq 
To find an explicit dependence of  $\bar\alpha$ on  $L$, one has to know the inverse function $F^{-1}[\alpha]$, which we know only perturbatively.

The beta function and, accordingly, the function $F[\alpha]$ is usually calculated within the perturbation theory, leaving the first few terms of expansion and then solving equation (\ref{eq}) exactly. In this case, the function $F^{-1}[\alpha]$ is easily obtained in one loop as a geometrical progression, more difficult in two loops as the Lambert W function, and  does not work at all in higher loops. Therefore, approximate formulas such as the inverse logarithm decomposition in terms of $1/\log(Q^2/\Lambda^2)$ are usually used.

It turns out that if we approach this problem differently and solve the renormalization group equation obtained by cutting the series not exactly but also approximately, we get very simple formulas for successive series of logarithmic expansion  (\ref{PT}) in terms of elementary functions containing only logarithms from more complex arguments.
At the same time, the differential RG equations  for all successive approximations turn out to be  linear and are easily integrated.

Indeed, consider the perturbative expansion of the running coupling as a series of logarithms
\beq
\bar\alpha[\alpha, L]=\alpha \left(1+\sum_{n=1}^\infty \alpha^n L^n  A_n 
+\sum_{n=2}^\infty \alpha^n L^{n-1}  B_n  
+\sum_{n=3}^\infty \alpha^n L^{n-2}  C_n  
+ \ldots\right) \label{PT}
\eeq
Here $A_n, B_n, C_n$  are some coefficients calculated within perturbation theory.
The first sum in eq.(\ref{PT}) corresponds to the leading log (LL) approximation, the next one to the NLL, then the NNLL, etc.

The advocated method of approximate solution of the RG equation is the following: we look for a solution to eq.(\ref{eq})  in the form of loop expansion
\beq
\bar\alpha[\alpha, L]= \alpha_1+\alpha_2+\alpha_3+...= \sum_{n=1}^\infty \alpha_n,
\eeq
where the functions  $\alpha_k[\alpha,L]$ sum up the infinite series of logarithms from (\ref{PT})  of the form $\sum_{n=1}^\infty \alpha^n L^{n-k+1}$,  respectively.  Namely, $\alpha_1$ sums all LLs 
$\sum_{n=1}^\infty \alpha^n L^n  A_n $, $\alpha_2$ sums  NLLs  $\sum_{n=1}^\infty \alpha^n L^n  B_n $, and so on. 

In this case, the function $\alpha_1$ obeys the one-loop equation
\beq
\frac{d \alpha_1}{dL}=\beta_0 \alpha_1^2, \ \ \ \ \alpha_1(\alpha,0)=\alpha,
\eeq
and all the other functions $\alpha_n, \ n>1$ obey the linearised equations obtained from  (\ref{eq}) keeping the terms of the same order of magnitude, and assuming that  $\alpha_n\sim \alpha_1^n$
\beqa
\frac{d \alpha_2}{dL}&=&2\beta_0 \alpha_1 \alpha_2+\beta_1 \alpha_1^3, \\
\frac{d \alpha_3}{dL}&=&2\beta_0 \alpha_1 \alpha_3+\beta_0 \alpha_2^2+3\beta_1 \alpha_1^2\alpha_2+\beta_2 \alpha_1^4, \\
\frac{d \alpha_4}{dL}&=&2\beta_0 \alpha_1 \alpha_4+2\beta_0 \alpha_2\alpha_3+3\beta_1 \alpha_1^2\alpha_3+3\beta_1 \alpha_1\alpha_2^2+4 \beta_2 \alpha_1^3\alpha_2+\beta_3\alpha_1^5, \\
\frac{d \alpha_5}{dL}&=&2\beta_0 \alpha_1 \alpha_5+2\beta_0 \alpha_2\alpha_4+\beta_0 \alpha_3^2+\beta_1 \alpha_2^3 + 3\beta_1 \alpha_1^2 \alpha_4+6\beta_1 \alpha_1\alpha_2\alpha_3 \nonumber \\ 
 &+ &4\beta_2 \alpha_1^3\alpha_3 +6\beta_2 \alpha_1^2 \alpha_2^2 + 5\beta_3 \alpha_1^4 \alpha_2 +\beta_4 \alpha_1^6, \\
& ... &\nonumber 
\eeqa
with the boundary conditions such as $\alpha_n[\alpha,0]=0, \ n>1$.

The solution of equation for  $\alpha_1$  has a form of a geometrical progression
\beq
\alpha_1[\alpha,L]=\frac{\alpha}{1-\beta_0\alpha L}, \label{prog}
\eeq
and solutions to the linear equations for the other functions can be easily obtained in a simple analytic form in terms of elementary functions
\beqa
\alpha_2[\alpha,L]&\!\!=\!\!&\bar \beta_1\alpha_1^2 \log\left(\frac{\alpha_1}{\alpha}\right), \label{2}\\
\alpha_3[\alpha,L]&\!\!=\!\!&\bar \beta_1^2 \alpha_1^3\left(\log^2\left(\frac{\alpha_1}{\alpha}\right)+\log\left(\frac{\alpha_1}{\alpha}\right)\right)+\alpha_1^2(\alpha_1-\alpha)(\bar \beta_2-\bar \beta_1^2), \\
\alpha_4[\alpha,L]&\!\!=\!\!&\bar \beta_1^3\alpha_1^4\left(\log^3\left(\frac{\alpha_1}{\alpha}\right)+\frac 52\log^2\left(\frac{\alpha_1}{\alpha}\right)\right)-\bar \beta_1\alpha_1^3(2\bar \beta_1^2(\alpha_1\!\!-\!\!\alpha)-\bar \beta_2(3\alpha_1\!\!-\!\!2\alpha))\log\left(\frac{\alpha_1}{\alpha}\right) \nonumber \\ &-&\frac 12 \bar \beta_1^3\alpha_1^2(\alpha_1-\alpha)^2-\bar \beta_1\alpha_1^2 \alpha(\alpha_1-\alpha)+\frac 12\bar \beta_3\alpha_1^2(\alpha_1^2-\alpha^2), \\
\alpha_5[\alpha,L]&\!\!=\!\!&
\alpha_1^5 \bar \beta_1^4\left(\log^4\left(\frac{\alpha_1}{\alpha}\right) + \frac{13}{3}\log^3\left(\frac{\alpha_1}{\alpha}\right)
   - \frac 32 \log^2\left(\frac{\alpha_1}{\alpha}\right) - 4 \log\left(\frac{\alpha_1}{\alpha}\right)\right) \nonumber\\ &+&
  6 \alpha_1^5  \bar \beta_2\bar \beta_1^2\log^2\left(\frac{\alpha_1}{\alpha}\right)  + 
  3 \alpha_1^4  \alpha  \bar \beta_1^4\log^2\left(\frac{\alpha_1}{\alpha}\right)- 
  3 \alpha_1^4 \alpha  \bar \beta_2\bar \beta_1^2 \log^2\left(\frac{\alpha_1}{\alpha}\right)\nonumber \\ &+& 
  5 \alpha_1^4 \alpha  \bar \beta_1^4\log\left(\frac{\alpha_1}{\alpha}\right)- 
  5 \alpha_1^4 \alpha \bar \beta_2 \bar \beta_1^2 \log\left(\frac{\alpha_1}{\alpha}\right) + 
  2 \alpha_1^5 \bar \beta_3 \bar \beta_1 \log\left(\frac{\alpha_1}{\alpha}\right)  - 
  \alpha_1^3 \alpha^2 \bar \beta_1^4 \log\left(\frac{\alpha_1}{\alpha}\right) \nonumber \\ &+&
  2 \alpha_1^3  \alpha^2 \bar \beta_2 \bar \beta_1^2\log\left(\frac{\alpha_1}{\alpha}\right)  - 
  \alpha_1^3 \alpha^2 \bar \beta_3\bar \beta_1 \log\left(\frac{\alpha_1}{\alpha}\right) + 
  3 \alpha_1^5 \bar \beta_2 \bar \beta_1^2 \log\left(\frac{\alpha_1}{\alpha}\right) \nonumber \\ &+& 
  1/6 (\alpha_1 - \alpha) \alpha_1^2 (7 \alpha_1^2 \bar \beta_1^4  - 5 \alpha_1 \alpha \bar \beta_1^4 - 
     18 \alpha_1^2  \bar \beta_2 \bar \beta_1^2 + 12 \alpha_1 \alpha  \bar \beta_2\bar \beta_1^2 \nonumber \\ &-&  
     \alpha_1^2  \bar \beta_3 \bar \beta_1 - \alpha_1 \alpha  \bar \beta_3 \bar \beta_1 + 10 \alpha_1^2  \bar \beta_2^2  - 
     8 \alpha_1 \alpha  \bar \beta_2^2+ 2 \alpha_1^2  \bar \beta_4 + 2 \alpha_1 \alpha  \bar \beta_4 \nonumber \\ &+&  
     6 \alpha^2  \bar \beta_2 \bar \beta_1^2  - 4 \alpha^2  \bar \beta_3 \bar \beta_1 - 2 \alpha^2  \bar \beta_2^2  + 
     2 \alpha^2  \bar \beta_4 - 2 \alpha^2   \bar \beta_1^4 ),
\label{4} \\
& ...,& \nonumber 
\eeqa
where $\alpha_1$ is given by (\ref{prog}) and we adopted the notation $\bar \beta_i=\beta_i/\beta_0$. Note that these solutions correspond to the exact sum of the logarithms in the $n$-th order.

Here it is appropriate to switch to the above-mentioned variable $\hat L \equiv \log\left(Q^2/\Lambda^2\right)$, where $\Lambda$ is related to $\mu$ as follows:
$$\log\left(\Lambda^2/\mu^2\right)=-\int \frac{dg}{\beta(g)}.$$
In the leading order, one has
\beq
\beta_0\log\left(\Lambda^2/\mu^2\right) =\frac{1}{\alpha}+\bar \beta_1 \log\alpha +(\bar \beta_2-\bar \beta_1^2) \alpha+ ...
\eeq
Then, equations (\ref{prog}, \ref{2}-\ref{4})  take the form
\beqa
\hat \alpha_1[\hat L]&=&\frac{1}{-\beta_0 \hat L}, \label{5}\\
\hat \alpha_2[\hat L]&\!\!=\!\!&\bar \beta_1 \hat \alpha_1^2 \log\left(\hat \alpha_1\right), \\
\hat \alpha_3[\hat L]&\!\!=\!\!&\bar \beta_1^2\hat \alpha_1^3\left(\log^2\left(\hat \alpha_1\right)+\log\left(\hat \alpha_1\right)\right)+\hat \alpha_1^3\left(\bar \beta_2\!-\!\bar \beta_1^2\right), \\
\hat \alpha_4[\hat L]&\!\!=\!\!&\bar \beta_1^3\hat \alpha_1^4(\log^3(\hat \alpha_1)+\frac 52\log^2(\hat \alpha_1))+\bar \beta_1\hat \alpha_1^4(3\bar \beta_2\!-\!2\bar \beta_1^2)\log(\hat \alpha_1)  +\hat \alpha_1^4 \frac 12(\bar \beta_3\!- \!\bar \beta_1^3),\\
\hat \alpha_5[\hat L]&\!\!=\!\!&\bar \beta_1^4\hat \alpha_1^5\left(\log^4\left(\hat \alpha_1\right)\!+\!\frac{13}{3} \log^3\left(\hat \alpha_1\right)\!-\!\frac 32\log^2\left(\hat \alpha_1\right)\!-\!4\log\left(\hat \alpha_1\right)\right)\!+\!6\bar \beta_2\bar \beta_1^2\hat \alpha_1^5\log^2(\hat \alpha_1) \nonumber \\
&+&(3\bar \beta_2\bar \beta_1^2\!+\!2\bar \beta_3 \bar \beta_1)\hat \alpha_1^5\log(\hat \alpha_1)\!+\!\hat \alpha_1^5(\frac 13 \bar \beta_4\!-\!\frac 16\bar \beta_3 \bar \beta_1\!+\!\frac 53\bar \beta_2^2\!-\!3\bar \beta_2\bar \beta_1^2\!+\!\frac 76 \bar \beta_1^4), \nonumber  \\
& ...& \label{6} 
\eeqa

Notice, that the subsequent functions $\hat \alpha_k[\hat L]$ are actually the functions of $\hat \alpha_1$,
namely $\hat \alpha_k[\hat \alpha_1]$.  It is convenient to use this form later.

The obtained solutions can be written in a more familiar form of inverse logarithm expansion (remind that
$\hat \alpha_1=1/(-\beta_0 \hat L)$)
\beqa
\alpha[\hat L]&=& \hat \alpha_1\left\{1+\bar \beta_1 \hat \alpha_1\log(\hat \alpha_1)+\hat \alpha_1^2(\bar \beta_1^2(\log^2(\hat \alpha_1)+\log(\hat \alpha_1)-1)+\bar \beta_2)\right.\nonumber\\
&+&\left. \hat \alpha_1^3 \left(\bar \beta_1^3(\log^3(\hat \alpha_1)+\frac 52\log^2(\hat \alpha_1)-2\log(\hat \alpha_1)-\frac 12)+3\bar \beta_1\bar \beta_2\log(\hat \alpha_1)+\frac 12\bar \beta_3\right) \right. \nonumber\\
&+& \left. \hat \alpha_1^4 \left( \bar \beta_1^4\left(\log^4(\hat \alpha_1)\!+\!\frac{13}{3} \log^3(\hat \alpha_1)\!-\!\frac 32\log^2(\hat \alpha_1)\!-\!4\log(\hat \alpha_1)+\frac 76 \right) \right.\right. \label{inv}\\
&+& \left. \left.\bar \beta_1^2\bar \beta_2(6\log^2(\hat \alpha_1)+3\log(\hat \alpha_1) -3)
+\bar \beta_1 \bar \beta_3 \left(2\log(\hat \alpha_1)-\frac 16\right)\!+\!\frac 13 \bar \beta_4+\frac 53\bar \beta_2^2 \right)+... \right\}\nonumber
\eeqa
Using our approach, we can easily continue this expansion to any desired order. It will contain only the logarithms and no special functions whatsoever. Similar solutions can be obtained iteratively from the renormalisation group equation, which is well established in the literature \cite{Shirkov:1981mb,ParticleDataGroup:2020ssz}.

\section{Vertical Summation and Nested Logs}

The next nontrivial step is to sum the main and subsequent logarithms
already in the obtained solutions, i.e. if earlier we summed the logarithms in (\ref{PT}) horizontally, now we sum them in equations (\ref{5}-\ref{6}) vertically.
The most amazing thing is that they are given by the same functions $\hat \alpha_n$, but with a different argument. We denote these new functions by $\hat \alpha_k^{(1)}$. In this notation, our old functions 
$\hat \alpha_k$ can be  called $\hat \alpha_k^{(0)}$.
For the principal logarithms we again have a geometrical progression
\beq
\hat \alpha_1^{(1)}=\frac{\hat \alpha_1}{1-\bar \beta_1\hat \alpha_1 \log(\hat \alpha_1)}, \label{v1}
\eeq
and for the subsequent terms, respectively
\beqa
\hat \alpha_2^{(1)}&=& \bar \beta_1( \hat \alpha_1^{(1)})^2 \log(\hat \alpha_1^{(1)}/\hat \alpha_1)=  \hat \alpha_1^2~\hat 
\alpha_2\left[\frac{\hat \alpha_1^{(1)}}{\hat \alpha_1}\right], \\
\hat \alpha_3^{(1)}&=& \hspace{2.6cm}
=\hat \alpha_1^3 \hat \alpha_3\left[\frac{\hat\alpha_1^{(1)}}{\hat \alpha_1}\right],\\
&&... \nonumber\\
\hat \alpha_n^{(1)}&=&\hspace{2.6cm}
= \hat \alpha_1^n \hat \alpha_n\left[\frac{\hat \alpha_1^{(1)}}{\hat \alpha_1}\right]. \label{vn}
\eeqa

Interestingly, the summation of the leading logarithms does not end up here.  The expressions for $\hat \alpha_i^{(1)}$ contain all the same logarithms but from more complex arguments, and they can also be summed, and the sums are given by the same functions $\hat \alpha_i$ but with different arguments
\beqa
\hat \alpha_1^{(2)}&=&\frac{\hat \alpha_1^{(1)}}{1-\bar \beta_1\hat \alpha_1^{(1)} \log(\hat \alpha_1^{(1)}/\hat \alpha_1)}, \\
\hat \alpha_2^{(2)}&=& (\hat \alpha_1^{(1)})^2~\hat \alpha_2\left[\frac{\hat \alpha_1^{(2)}}{\hat \alpha_1^{(1)}}\right], \\
\hat \alpha_3^{(2)}&=& 
(\hat \alpha_1^{(1)})^3 ~\hat \alpha_3\left[\frac{\hat  \alpha_1^{(2)}}{\hat \alpha_1^{(1)}}\right],\\
&&... \nonumber\\
\hat \alpha_n^{(2)}&
= &(\hat \alpha_1^{(1)})^n ~\hat \alpha_n\left[\frac{\hat \alpha_1^{(2)}}{\hat \alpha_1^{(1)}}\right] .\label{vn}
\eeqa

Natural generalisation of these formulas for further steps is given by the following  genuine expression:
\beqa
\hat \alpha_1^{(m)}&=&\frac{\hat\alpha_1^{(m-1)}}{1-\bar \beta_1\hat \alpha_1^{(m-1)}\log(\hat \alpha_1^{(m-1)}/\hat \alpha_1^{(m-2)})}, \ \hat \alpha_1^{(0)}=\hat \alpha_1, \hat \alpha_1^{(-1)}=1,
\label{1m}\\
\hat \alpha_n^{(m)}&
=& (\hat \alpha_1^{(m-1)})^n ~~\hat \alpha_n\left[\frac{\hat \alpha_1^{(m)}}{\hat \alpha_1^{(m-1)}}\right], \ n\geq 2, m\geq 1 \label{nm}.
\eeqa
These formulas allow for a further improvement of the approximation due to further summation of vertical infinite series, and this procedure continues endlessly. At each step, one has the same functions but with different arguments, and one can cut this process at a point when a new term of the beta-function appears. For the calculated N-loop beta-function, one has  for the best approximation
\beq
\bar \alpha=\sum_{n=1}^{N}\hat \alpha_n^{(N-1)},
\eeq
where $\hat \alpha_n^{(N)}$ are given by eqs.(\ref{nm},\ref{1m}). For example, in three loops  we get the best approximation as
\beqa
\bar\alpha_{1+2+3}= \sum_{k=1}^3\hat \alpha_k^{(2)}=
\frac{\hat \alpha_1^{(1)}}{1-\bar\beta_1\hat \alpha_1^{(1)} \log\left(\frac{\hat \alpha_1^{(1)}}{\hat \alpha_1}\right)}+(\hat \alpha_1^{(1)})^2~\hat \alpha_2\left[\frac{\hat \alpha_1^{(2)}}{\hat \alpha_1^{(1)}}\right]+
(\hat \alpha_1^{(1)})^3~ \hat \alpha_3\left[\frac{\hat  \alpha_1^{(2)}}{\hat \alpha_1^{(1)}}\right],
\eeqa
with
\beqa
\hat \alpha_1[\hat L]&=&1/(-\beta_0 \hat L), \\ \hat \alpha_1^{(1)}[\hat\alpha_1]&=&\frac{\hat \alpha_1}{1-\bar \beta_1\hat \alpha_1 \log(\hat \alpha_1)},\\
\hat \alpha_2[x]&=&\bar \beta_1 x^2\log(x),\\
\hat \alpha_3[x]&=& \bar \beta_1^2x^3(\log^2(x)+\log(x)-1)+\bar \beta_2 x^3.
\eeqa

\begin{figure}[htbp]
\begin{minipage}{0.45\textwidth}
\includegraphics[width=\linewidth]{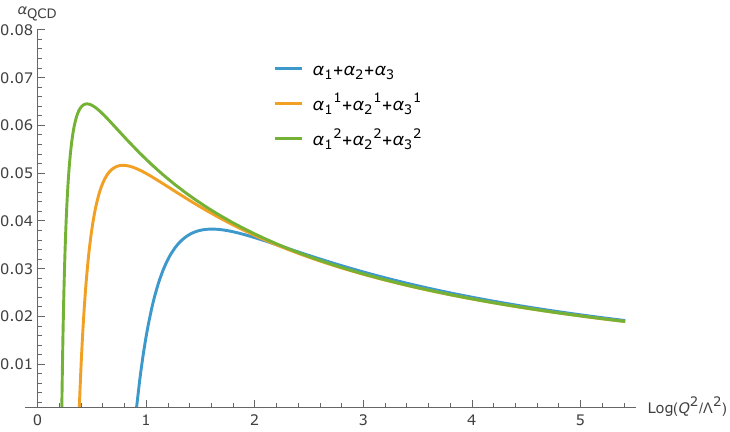}
\caption{Comparison of the three-loop running coupling $\bar \alpha_s(Q^2)$ in the first, second and  third 'nested' approximations.}\label{plots1}
\end{minipage}
\hfill
\begin{minipage}{0.45\textwidth}
 \includegraphics[width=0.9\linewidth]{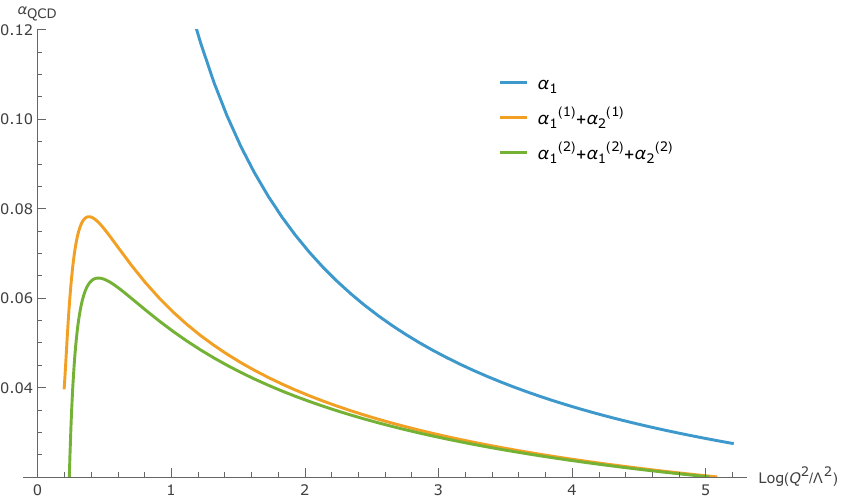}
\caption{Comparison of the one-loop and improved two-loop and three-loop running coupling $\bar \alpha_s(Q^2)$.}\label{plots2}
\end{minipage}
\end{figure}

For illustration, we calculate the running coupling $\bar \alpha$ in QCD in three loops  for  various approximations. We use here the three-loop beta-function in the $\bar{\text{MS}}$ scheme \cite{Tarasov:1980au,Larin:1993tp}
\beqa
\beta_0&=&-(11-2/3 n_f), \\
\beta_1&=&-(102-38/3 n_f),\\
\beta_2&=&-(2857/2-5033/18 n_f+325/54 n_f^2).
\eeqa
For 6 quark flavours this gives $\beta_0= -7, \beta_1=-26, \beta_2=65/2$. The  corresponding curves are shown in Figs.\ref{plots1} and \ref{plots2}

One can see that taking account of the subsequent approximations stabilises the curve in the low $Q^2$ region and leads to   smooth behaviour for high $Q^2$.

\section{Solution for the Green Functions} 

We now repeat the same procedure for an object with an anomalous dimension such as an amplitude, Green's functions, structure functions, potential, etc.
Then,  in a renormalizable theory, the RG equation takes the form \cite{Ovsyannikov:1956fa,Callan:1970yg,Symanzik:1970rt}
\beq
\left(\mu^2\frac{d}{d\mu^2}+\beta(\alpha)\frac{d}{d\alpha}-\gamma(\alpha)\right)\Gamma\left[\frac{Q^2}{\mu^2},\alpha\right]=0,
\eeq
where $Q^2$ is some characteristic variable such as a momentum or a field.
 This equation can also be rewritten as an ordinary differential equation
 \beq
 \frac{d \log\Gamma}{d\log(\mu^2)}=\gamma(\alpha).
 \eeq
 Additionally, there is a normalization condition $\Gamma[1,\alpha]=\Gamma_0\cdot(1+O(\alpha))$, which depends on the subtraction scheme.

 The last equation has the same form as the equation for the running coupling (\ref{eq}), and the solution can be  formally written as
 \beq
 \log(\Gamma)=\int_{\alpha}^{\bar \alpha}\frac{\gamma(x)}{\beta(x)}dx. \label{sol}
 \eeq
 Here the anomalous dimension and the beta function are given by perturbative expansion
 \beq
 \gamma(\alpha)=\gamma_0\alpha+\gamma_1\alpha^2+\gamma_2\alpha^3+...
 \eeq
 Now following our procedure for allocating the leading, nex-to-leading, and so on logarithms, we rewrite $\log(\Gamma)$ as
 \beq
 \log(\Gamma)=\log(\Gamma_1)+\log(\Gamma_2)+\log(\Gamma_3)+ ...,
\eeq
where $\log(\Gamma_1)$ corresponds to the contribution of the leading logarithms, $\log\Gamma_2$- next-to-leading ones, and so on.
 
 Then eq.(\ref{sol}) can be written as
 \beq
 \log(\Gamma_1)+\log(\Gamma_2)+\log(\Gamma_3)+ ....=\int_{\alpha}^{\alpha_1+\alpha_2+\alpha_3+...}\frac{\gamma_0}{\beta_0 x}(1+c_1 x+c_2 x^2+...)dx, 
 \eeq
 where
 $$ c_1=\bar \gamma_1 -\bar \beta_1, \ \ \  c_2=\bar \gamma_2 -\bar \beta_2-\bar \gamma_1 \bar \beta_1+\bar \beta_1^2,  \ \ \ ...,$$
where we use the ntation $\bar \gamma_i=\gamma_i/\gamma_0$.

 Selecting  appropriate orders of magnitude, we obtain  the following expressions for the subsequent terms of the $\Gamma$-function expansion:
 \beqa
 \log\Gamma_1&=&\frac{\gamma_0}{\beta_0}\log\frac{\alpha_1}{\alpha}, \\
 \log\Gamma_2&=&\frac{\gamma_0}{\beta_0}\left(\frac{\alpha_2}{\alpha_1}+c_1\alpha_1\right), \\
  \log\Gamma_3&=&\frac{\gamma_0}{\beta_0}\left(\frac{\alpha_3}{\alpha_1}+c_1\alpha_2+c_2\alpha_1^2\right),\\
  &+& ... 
\eeqa 
Combining everything together, we get
\beq
\Gamma=\Gamma_0 \cdot \left(\frac{\alpha_1}{\alpha}\right)^{\gamma_0/\beta_0}e^{\frac{\gamma_0}{\beta_0}\left(\left(\frac{\alpha_2}{\alpha_1}+c_1\alpha_1\right)+\left(\frac{\alpha_3}{\alpha_1}+c_1\alpha_2+c_2\alpha_1^2\right)+...\right)}, \label{solg}
\eeq
where the functions $\alpha_n$ are given by the above solutions (\ref{2}-\ref{4}). 
This expression can be expanded over the inverse logarithms like the running coupling $\alpha(\hat L)$ (\ref{inv})
\beq
\Gamma={\Gamma_0} \cdot \left(\frac{1}{-\beta_0 \hat L}\right)^{\gamma_0/\beta_0}\left(1+\frac{\gamma_0\bar \beta_1}{\beta_0}\frac{\log \hat L}{-\beta_0 \hat L}+O(1/\hat L^2)\right).
\eeq
One can also switch in (\ref{solg}) from $\alpha_k$ to the improved solutions for the running coupling $\hat \alpha_k$ and further to $\hat \alpha_k^{(m)}$ and get the improved expansion.
The  calculated N orders of expansion of the $\gamma$ and $\beta$ functions, one should take $\alpha_k^{(N-1)}$.
  
 \section{Conclusion}
 
 To conclude, we have demonstrated that taking the right strategy, one can get relatively simple explicit formulas for the running coupling expansion in the asymptotic regime. These solutions contain only logarithms and no special functions and correspond to the summation of an infinite series of leading, next-to-leading, and so on logarithms.  
 
 This method can  also be applied to  objects with anomalous dimensions such as the Green function or  structure functions typical of QCD. The obtained explicit formulas directly reproduce the inverse logarithm expansion popular in QCD to any given order but provide better approximation. The possibilities of applying the obtained results to analytical perturbation theory also look promising\cite{Shirkov:1997wi}. 
 
 \section*{Acknowledgments}
 The authors are grateful to S.V. Mikhailov for useful comments and discussions.
 
\bibliography{refs}
\bibliographystyle{unsrt} 

\end{document}